\documentclass[cits]{PoS}
\usepackage{amsmath}

\title{How much of the inflaton potential do we see?}

\ShortTitle{How much of the inflaton potential do we see?}

\author{\speaker{Wessel Valkenburg}
\\
LAPTH\footnote{Laboratoire de Physique
Th\'eorique d'Annecy-le-Vieux, UMR5108}, Universit\'e de Savoie \&
CNRS\\
9 chemin de Bellevue, BP110\\
F-74941 Annecy-le-Vieux Cedex\\
France\\
E-mail:\email{wessel.valkenburg@lapp.in2p3.fr}}
\date{\today}
\FullConference{Carg\`ese Summer School: Cosmology and Particle Physics Beyond the Standard Models\\
                July 30 - August 11, 2007\\
                Institut d'Etudes Scientifiques de Carg\`ese}

              \abstract{We discuss the latest constraints on a
                Taylor-expanded scalar inflaton potential, obtained
                focusing on its observable part only. This is in
                contrast with other works in which an extrapolation of
                the potential is applied using the slow-roll
                hierarchy. We find significant differences. The
                results discussed here apply to a broader range of
                models, since no assumption about the invisible
                e-folds of inflation has to be made, thereby remaining
                conservative.}

\begin{document}


\maketitle

\section{Introduction}
The cosmology of the early universe is hindered by the question about
initial conditions. By inflating a region, so small that it does not
sense a non flat universe, up to proportions at least larger than the
presently observable universe, the paradigm of cosmic inflation wipes
out any other initial conditions than an approximately spatially flat,
homogeneous and isotropic universe~\cite{Starobinsky:1980te}. At the
same time it provides a mechanism for the generation of correlations
in perturbations on all scales, observed today in the cosmic microwave
background (CMB) and the large scale structure
(LSS)~\cite{Starobinsky:1979ty}.

In order to create the coherence on scales as large as the Hubble
horizon today, inflation must last about $30 - 60$ e-folds after the
presently observed spectrum has left the Hubble horizon. The shape of
the spectrum of these observed perturbations presently gives us the
only hint on the physics that can have driven inflation.  The
observations alone though, give no information on inflation after the
creation of the observed spectrum.  Many previous works have used the
formalism of flow equations~\cite{Steinhardt:1984jj} to extrapolate
beyond the observed primordial power spectrum, in order to select only
those spectra that correspond to a scalar-field potential that drives
inflation long
enough~\cite{Spergel:2006hy,Peiris:2006ug,Easther:2006tv}. This
extrapolation puts strong bounds on the shape of the
potentials~\cite{Easther:2006tv}. However, no work has focused on the
shape of the potential within only the observable window, using
present-day data.

In a recent work we presented new bounds on the shape of the inflaton
potential within the observable window only, thereby remaining
conservative about what (humble or exotic) mechanism drives inflation
in the subsequent e-folds~\cite{Lesgourgues:2007gp}. The results were
obtained by matching a Taylor-expanded smooth scalar-field potential
directly to the data numerically (the data being
WMAP\footnote{Wilkinson Microwave Anisotropy
Probe~\cite{Spergel:2006hy,Page:2006hz}} 3-year data and the
SDSS-LRG\footnote{Sloan Digital Sky Survey - Luminous Red Galaxy, data
release 4~\cite{Tegmark:2006az}} sample). These results apply to any
theory, that within the observable window effectively has one scalar
degree of freedom (one inflaton) with a smooth potential.

\section{Conservative approach}
In our calculations, we evolved the universe for a given $V(\phi)$.
During a stage of scalar field domination, the Friedmann equation and
the equation of motion of the scalar field can be written with the
usual quantities as
  \begin{align}
    \dot \phi &= -\frac{m_P^2}{4 \pi} H'(\phi),\label{eq:eom}\\
    \left[H'(\phi)\right]^2-\frac{12 \pi}{m_P^2}H^2(\phi) &= -\frac{32 \pi^2}{m_P^4}V(\phi),\label{eq:friedmann}
  \end{align}
  where $\dot{\phantom{.}}$ denotes a time derivative and $'$ denotes
  a derivative with respect to $\phi$. Initial conditions for the
  evolution can be chosen in terms of $V(\phi)$ and $\dot \phi$. By
  demanding that the attractor solution for $\dot \phi$ is reached
  just before the observable scales leave the horizon
, we chose to decrease the
  number of free parameters.
%
The attractor solution is the solution
 in which the accelerating force on the inflaton and the Hubble
 friction precisely cancel. In practice this implies that inflation
 starts {\em at least} a few e-folds before observable modes leave the
 horizon.
\begin{table}[t]
\begin{center}
\begin{tabular}{l|ccc}
Parameter & $v=2$ & $v=3$ & $v=4$ \\
\hline
\hline
$\Omega_b h^2$ & 
$0.022 \pm 0.001$ & $0.022 \pm 0.001$ & $0.022 \pm 0.001$ \\
$\Omega_{cdm} h^2$ & 
$0.109 \pm 0.004$ & $0.109 \pm 0.004$ & $0.109 \pm 0.004$ \\
$\theta$ & 
$1.041 \pm 0.004$ & $1.041 \pm 0.004$ & $1.040 \pm 0.004$ \\
$\tau$ & 
$0.08 \pm 0.03$ & $0.09 \pm 0.03$ & $0.10 \pm 0.03$ \\
$\ln \! \left[\frac{128 \pi 10^{10} V^3_*}{3 V'^2_* m_P^6}\right]$ &
 $3.06 \pm 0.06$ & $3.07 \pm 0.06$ & $3.11 \pm 0.08$ \\
$\left(\frac{V'_*}{V_*}\right)^2 \!\!\! m_P^2$ & 
$<0.4$ & $<0.4$ & $<0.8$ \\
$\frac{V''_*}{V_*} m_P^2$ & 
$0.1 \pm 0.5$ & $-0.2 \pm 0.6$ & $0.4 \pm 0.9$ \\
$\frac{V'''_*}{V_*} \frac{V'_*}{V_*} m_P^4$ & 
0 & $8 \pm 5$ & $13 \pm 11$ \\
$\frac{V''''_*}{V_*} \left(\frac{V'_*}{V_*} \right)^2 \!\!\! m_P^6$ & 
0 & 0 & $200 \pm 150$ \\
\hline
$- \ln {\cal L}_{\rm max}$ & 2688.3 & 2687.2 & 2687.2 \\
\end{tabular}
\end{center}
\caption{Bayesian 68\% confidence limits for
$\Lambda$CDM inflationary models with a Taylor expansion of the
inflaton potential at order $v=2,3,4$ (with the primordial spectra
computed numerically). The last line shows the maximum
likelihood value. }\label{table:results} 
\end{table}
To do a Monte-Carlo simulation we used the publicly available code
{\sc Cosmomc}~\cite{Lewis:2002ah}, which we extended with our own
available module~\cite{Lesgourgues:2007gp} to numerically calculate
the primordial power spectrum (including tensors), given some
potential. We assumed a spatially flat standard $\Lambda$CDM-model,
with the free parameters shown in Table~\ref{table:results}.  The
motivation to choose a Taylor expansion in stead of a hierarchy of
slow-roll parameters, is that we are interested in information on the
observable potential only. Both parametrizations are infinite
dimensional, such that a mapping from one to another is only bijective
if one knows the vectors in infinite dimensions or when both basis are
simply the same. The slow-roll hierarchy is useful for extrapolating,
where the Taylor expansion is more practical within a strict
window. The last statement remains a matter of taste, given that the
number of free parameters remains the same.

\section{Current bounds}
The new bounds on the derivatives of the inflaton potential, are
displayed in Table~\ref{table:results}. The number of degrees of
freedom in the potential is denoted by the parameter $v$, where $v=2$
denotes an expansion up to the second derivative, $v=3$ up to $V'''$,
etc.  Central values and error bounds are obtained by marginalization
over all other parameters and given at $68\%$ confidence level.  All
solutions are symmetric under a sign change in $V'$ (the inflaton
always rolls down) and $V'''$.
\begin{figure}[t]
\begin{center}
\includegraphics[height=.45\textwidth]{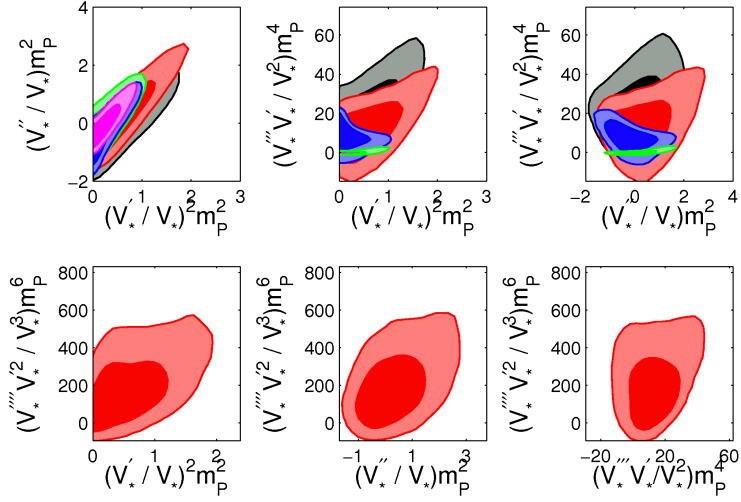}
\end{center}
\caption{\label{fig:triV} Two-dimensional 68\% and 95\% confidence level
contours based on WMAP 3-year and the SDSS LRG spectrum, for
the parameters describing the inflaton potential,
obtained directly from the MCMC in the case of models 
$v=2$ (magenta), $v=3$ (blue), $v=4$ (red), or derived
from second-order formulas for models $p=2$ (green), $p=3$ (black).}
\end{figure}

The best likelihoods of the simulations indicate that the improvement
by adding the fourth derivative is negligible. As could be expected,
the bounds on lower derivatives loosen when one allows higher
derivatives. 
In Figure~\ref{fig:triV} 
all 2D-correlations between the potential derivatives are given,
marginalized over all other parameters. In this figure results are
compared with those obtained by inverting the slow-roll approximation,
when probing a scalar and tensor primordial spectrum with a scalar
tilt ($p=2$) or an additional running of the tilt ($p=3$). The covered
area in parameter space is equally large in both approaches, however
the overlap is not $100\%$, 
which shows that indeed information is lost when mapping from one
basis to the other. This indicates that up to second order, the slow
roll approximation is not completely accurate within the observed
window (see also~\cite{Powell:2007gu}).  The primordial spectra of
both simulations on the other hand, which do not depend on
approximations, do fully overlap~\cite{Lesgourgues:2007gp}.
\begin{figure}[t]
\includegraphics[angle=-90,width=.5\textwidth]{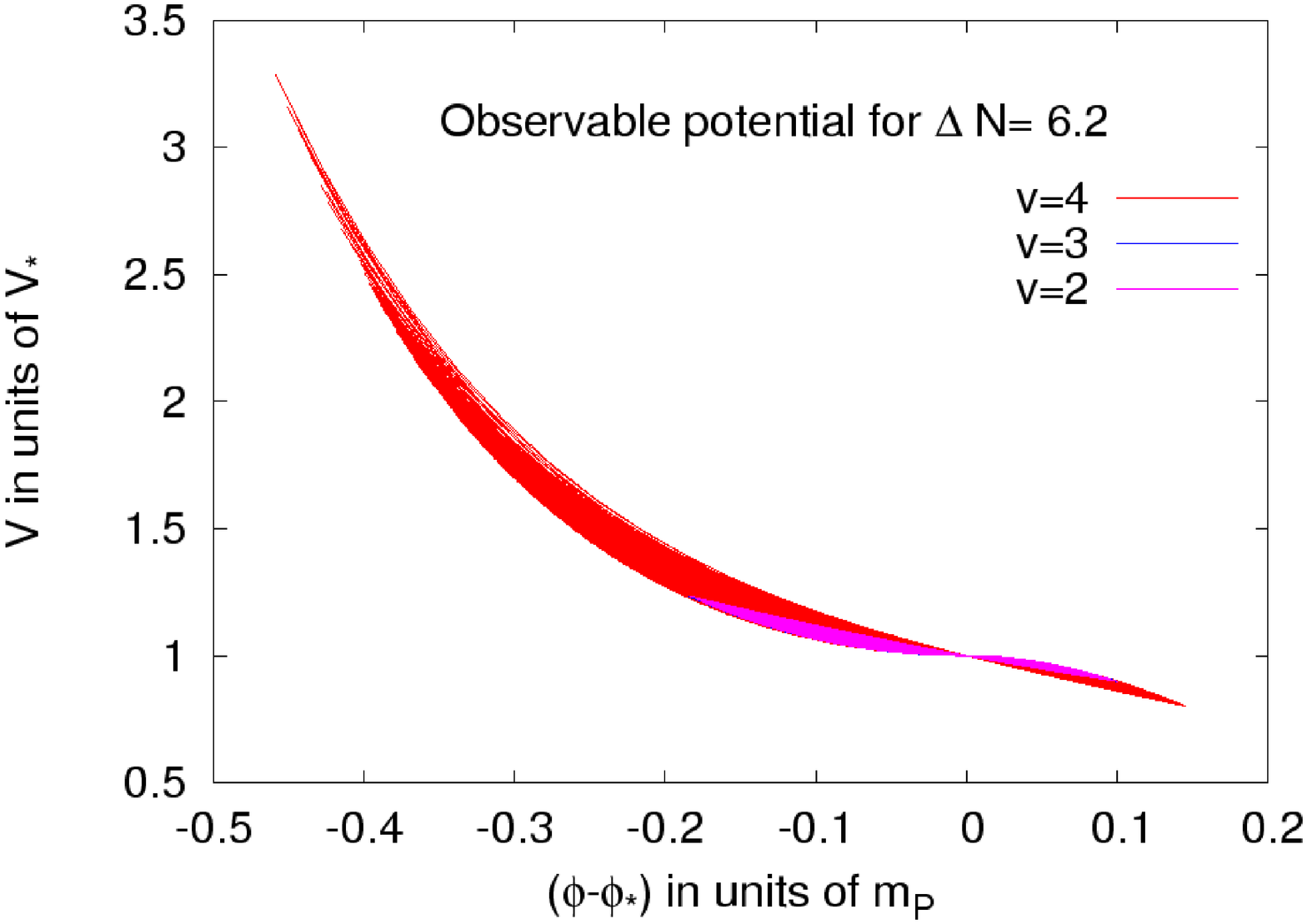}
\includegraphics[angle=-90,width=.5\textwidth]{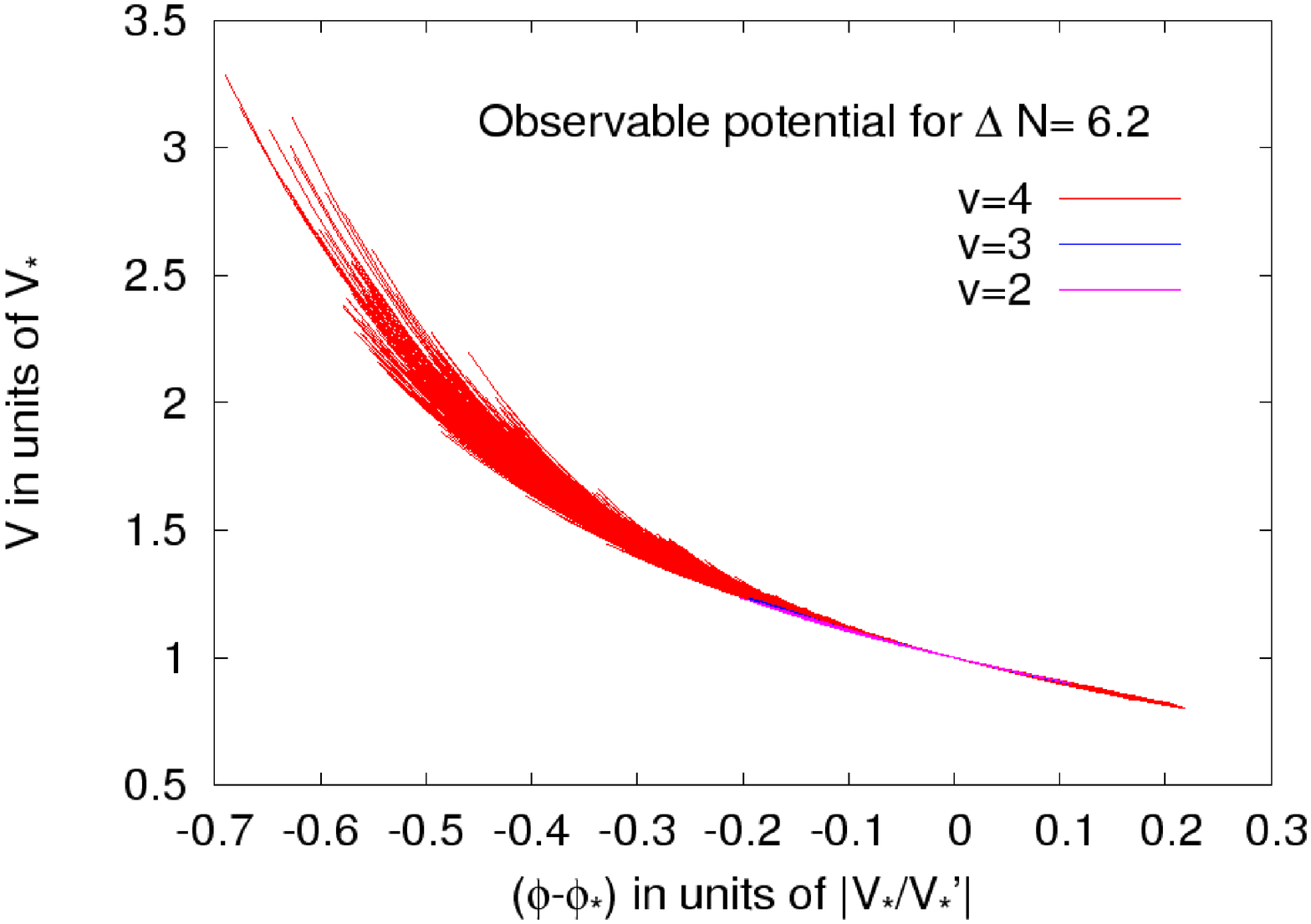}
\caption{{\em Left:} The shape of the scalar potential at the
$95\%$-confidence level, normalized to the value of the potentials at
the pivot scale. {\em Right:} The same, in a normalization such that
all first derivatives are equal, and focus can be laid on the higher
derivatives. }\label{fig:pot}
\end{figure}
Finally, in Figure~\ref{fig:pot} the inflaton potential as bounded by
present observations up to $95\%$ confidence level is shown, allowing
up to second, third and fourth derivative.

\section{Discussion}
Previous works relied on an extrapolation of data over a range which
is an order of magnitude broader than the range of the data itself,
which is fine from one point of view. Focusing on the data only, these
are the first results to constrain to potential up to such a high
order.

In equations~(\ref{eq:eom},\ref{eq:friedmann}), one
sees that $V(\phi)$ does not unambiguously 
define $H(\phi)$, however any $H(\phi)$ would uniquely define
$V(\phi)$. Hence the defining quantity should be $H(\phi)$, which will be
subject to a future work~\cite{future:slv}.

\section*{Acknowledgements}
The author would like to thank the organisers of the Carg\`ese Summerschool on Cosmology and Particles Beyond The Standard Models for an excellent stay.
This work was supported by the EU 6th
Framework Marie Curie Research and Training network ``UniverseNet''
(MRTN-CT-2006-035863).

\bibliography{refs}
\bibliographystyle{unsrt-wv}
\end{document}